\documentclass[reqno,11pt]{amsart}

\usepackage{amssymb,amsthm}
\usepackage{amsmath}
\usepackage{amsfonts}
\usepackage{dsfont}

\usepackage[a4paper,  margin=3.3cm]{geometry}
\usepackage[active]{srcltx}
\makeatletter\@addtoreset{equation}{section}\makeatother

\newtheorem{theorem}{Theorem}[section]

\newtheorem{proposition}[theorem]{Proposition}

\numberwithin{equation}{section}

\title[The Lee-Yang property of the BEG model]{The Lee-Yang property of the Blume-Emery-Griffiths model}

\author{ Yuri  Kozitsky}

\address{Instytut Informatyki i Matematyki, Uniwersytet Marii Curie-Sk{\l}odowskiej, plac Marii Curie-sk{\l}odowskiej 1, 20-031 Lublin, Poland; fax +48 81 537 5471}
\email{jurij.kozicki@mail.umcs.pl}

\begin{document}


\begin{abstract}
 The Lieb-Sokal theorem provides a tool for establishing the Lee-Yang property of the partition function of a spin model if the corresponding single-spin partition function has this property and the interaction term in its Hamiltonian is of the form $-\sum J_{ij} S_iS_j$ with $J_{ij}\geq 0$. According to the result of this paper, the spin $S=1$ model with the Hamiltonian $-\sum (J_{ij} S_iS_j + K_{ij}S_i^2S_j^2) + \sum \Delta S_i^2$ has the Lee-Yang property for all $K_{ij}\in \mathds{R}$ and $\Delta >0$, whenever each $J_{ij}$ exceeds an explicitly computed bound $C(\beta, \Delta, K_{ij})>0$, such that $C(\beta, \Delta, K_{ij})\to + \infty$ as $\beta \to 0$. The latter suggests that the Lee-Yang property of this model may hold only for sufficiently low temperatures.
\end{abstract}

\maketitle

\section{Introduction}


The celebrated Lee-Yang theorem states that the continuation to complex values of the external magnetic field $h$ of the partition function $Z^\beta_N$ of the collection of $N$ interacting Ising spins has purely imaginary zeros if the interaction is ferromagnetic; see \cite{FrR,LiebS,Newman,Ruelle} and the literature quoted therein. As a Laurent polynomial in $z=e^{\beta h}$ ($\beta$ being the inverse temperature), symmetric with respect to the change $z\to z^{-1}$, $Z^\beta_N$ is an even exponential-type entire function. Combining these facts, cf. \cite{Levin}, one readily 
concludes that it can be written in the form 
\begin{equation}
 \label{1}
 Z^\beta_N (h) = Z^\beta(0) \prod_{j=1}^\infty (1+ \gamma_j h^2),    
\end{equation}
where the collection $\{\gamma_j\}$ -- in this case, $\beta$- and $N$-dependent parameters --
satisfies
\begin{equation}
 \label{2}
 0 < \gamma_{j+1} \leq \gamma_j,  \qquad \sum_{ j=1}^\infty \gamma_j < \infty.
\end{equation}
Consequently, $Z^\beta_N$ is said to have the Lee-Yang property. 
With increasing $N$,
 the `first Lee-Yang zero' $i / \sqrt{\gamma_1}$ moves towards the origin; see \cite{JN1}.
In the limit $N\to \infty$, it may reach the origin and thus produce a singularity of the free energy density at $h=0$.
In view of this, the Lee-Yang theorem contributes to the collection of powerful methods for studying phase transitions in Ising and similar spin models. 

Obviously, \eqref{1} holds for $N=1$, where $Z^\beta_1 (h) = 2\cosh \beta h$. A natural generalization of the Lee-Yang theorem might be to claim the validity of \eqref{1} and \eqref{2} in the case of more general single-spin distributions that possess the Lee-Yang property, including those for `unbounded' spins, which are finite measures on the real line. 
 Newman \cite{Newman}, and then Lieb and Sokal \cite{LiebS}, did this. The key aspect of the latter work is the following representation
\begin{eqnarray}
 \label{3}
Z^\beta_N( h) & = & \int_{\mathds{R}^N}  \exp\left( -\beta H_N (S,h)\right)
\prod_{i=1}^N \chi(dS_i) \\ \nonumber
& = & \int_{\mathds{R}^N} \exp\left(  \sum_{1\leq i< j\leq N} \beta J_{ij} S_i S_j + \sum_{i=1}^N \beta h S_i \right) \prod_{i=1}^N \chi(dS_i) \\ \nonumber 
& = &  \left[ \exp\left(  \sum_{{1\leq i< j\leq N}} \beta J_{ij} D_i D_j) \right) \prod_{i=1}^N \phi(x_i) \right]_{{\rm all} \ x_i = \beta h}, \quad D_i = \frac{\partial}{\partial x_i},
\end{eqnarray}
where $\phi$ is related to the aforementioned finite measure $\chi$ by
\begin{equation}
 \label{4}
 \phi(x) = \int_{\mathds{R}} e^{x\sigma} \chi (d\sigma).
\end{equation}
The infinite-order differential operator in the last line of \eqref{3} can be defined for all $J_{ij} \geq 0$ if  $\phi$ is an entire function of order less than two, or of order two and of minimal type. The Lee-Yang property of $\chi$ means that $\phi$ can be written in the form 
\begin{equation}
 \label{5}
 \phi(x) = \hat{\phi}(x^2) = \phi(0)\prod_{j=1}^\infty (1+\gamma_j x^2), 
\end{equation}
where the collection $\{\gamma_j\}$ satisfies \eqref{2}, and hence $\hat{\phi}$ is a Laguerre entire function; see \cite{Fisk,Iliev} and also \cite{Koz}, where examples of single-spin measures possessing the Lee-Yang property are given. 

According to \cite{LiebS}, $Z^\beta_N$ given in \eqref{3} has the Lee-Yang property for all $\beta>0$ whenever $J_{ij} \geq 0$ for all $i,j$, and $\phi$ is as in \eqref{5}. The latter is a sufficient condition; it leaves open the possibility that $Z^\beta_N$ has the Lee-Yang property for $N\geq 2$, at least for some $\beta$, even if this condition is not satisfied. In \cite{KH}, such a possibility was shown to occur for the Blume-Capel model. Here we extend this result to the Blume-Emery-Griffiths (BEG) model; see \cite{BEG,Nihat,Lima,LimaM,Wu} for more details on this model, its versions and applications. 

The prototypical model that we study corresponds to the following partition function
\begin{equation}
\label{Ex}
\mathcal{Z}_N(x) =\int_{\mathds{R}^N} \exp\left(\sum_{{1\leq i< j\leq N}} [\mathcal{J}_{ij} S_i S_j + \mathcal{K}_{ij} S^2_i S^2_j ] + \sum_{i=1}^N x S_i \right)  \prod_{i=1}^N \chi_\theta (dS_i),
\end{equation}
with
\begin{equation}
\label{Ex1}
 \chi_\theta =  2\theta\delta_0 + \delta_{-1} + \delta_1,   \quad \theta >0,  
\end{equation}
where $\delta_a$ is Dirac's measure centered at $a\in \mathds{R}$. This model is related to the BEG model; see the next section. Note that the function corresponding (as in \eqref{4}) to $\chi_\theta$ given in \eqref{Ex1} is
\begin{equation}
 \label{Ex2}
 \phi_\theta(x) =  2 (\theta + \cosh x).
 \end{equation}
For $\theta>1$, it does not vanish on the imaginary axis, where $|\cosh x|\leq 1$. Then $\prod_{j=1}^N \phi_\theta (x_j)$ does not satisfy the sufficient condition mentioned above. Another essential obstacle to proving the Lee-Yang property of \eqref{Ex} is the quadrupole interaction term $\sum \mathcal{K}_{ij} S^2_i S^2_j$, which is incompatible with the technique of \cite{LiebS} or any other theory of this kind. In this paper, we prove that $\mathcal{Z}_N$ given in \eqref{Ex} has the Lee-Yang property for all $\theta>0$ and $\mathcal{K}_{ij}\in \mathds{R}$ if the interaction intensities $\mathcal{J}_{ij}\geq 0$ are large enough. To the best of our knowledge, this is the first result of this kind for the Blume-Emery-Griffiths model.

\section{The Result}
We begin by preparing the basic tool that we use to prove the main statement. 

\subsection{The Lieb-Sokal theorem}

First, we formulate the relevant
results of \cite{LiebS} in the form adapted to the context. To do this, we will need \emph{stable} polynomials. For a given $N\in \mathds{N}$, a complex polynomial,  $P$, is said to be stable if $P(z_1, \dots, z_N)\neq 0$ whenever $\Re z_j >0$ for all $j=1, \dots, N$; see \cite{Fisk,Ruelle} and the literature quoted in these publications. Let $\mathcal{F}_N$ be the set of exponential-type entire functions defined on $\mathds{C}^N$, which are: (a) not identically zero; (b) obtained as limits of the sequences of stable polynomials taken in the topology of uniform convergence on compact subsets of $\mathds{C}^N$. Clearly, the point-wise product $F_1 F_2$ is in $\mathcal{F}_N$ if both $F_1$ and $F_2$ are there. By Hurwitz' theorem, it follows that each $F\in \mathcal{F}_N$ verifies $F(z_1, \dots, z_N)\neq 0$ whenever $\Re z_j >0$ for all $j=1, \dots, N$. At the same time, by the Laguerre-P\'olya theorem, an even exponential-type entire function that does not vanish at zero is in $\mathcal{F}_1$ if and only if it is as in  \eqref{5}; see \cite[pages 8--23]{Iliev}. If $F\in \mathcal{F}_N$ is even, i.e., $F(x_1, \dots , x_N) = F(-x_1, \dots , -x_N)$, and $F(0, \dots , 0) \neq 0$, then $F(x, \dots , x)$ is as in \eqref{5}.       

For $G\in \mathcal{F}_N$, define
\begin{eqnarray}
  \label{5a}
 F(x_1, \dots , x_N) & = & \exp\left(\sum_{{1\leq i< j\leq N}} A_{ij} D_i D_j \right) G(x_1 , \dots, x_N), \quad  D_i = \frac{\partial}{\partial x_i}, \\ \nonumber \widehat{F}(x_1, \dots , x_N) & = & \prod_{1\leq i< j \leq N} (1+ A_{ij} D_iD_j) G(x_1 , \dots, x_N).
\end{eqnarray}
\begin{proposition}\cite[Propositions 2.6 and 2.7, page 161]{LiebS}
 \label{0pn}
 If $A_{ij}\geq 0$ for all  $1\leq i < j \leq N$, then for each $G\in \mathcal{F}_N$, both functions $F$ and $\widehat{F}$ given in \eqref{5a}
are in $\mathcal{F}_N$.
\end{proposition}
For $\theta \leq 1$, by solving $\phi_\theta(x) =0$, see \eqref{Ex2}, we obtain $e^x = \theta \pm i \sqrt{1-\theta^2}$. Therefore, $\phi_\theta$ is an exponential-type entire function of the form \eqref{5}, i.e., $\phi_\theta\in \mathcal{F}_1$. Then $\prod_{j=1}^N \phi_\theta (x_j)$ is in $\mathcal{F}_N$, and the direct corollary of Proposition \ref{0pn} is the following statement.
\begin{proposition} \cite[Corollary 3.3, page 165]{LiebS}
 \label{1pn} 
Let $\phi_\theta$ in \eqref{Ex2} correspond to $\theta\leq 1$, $\mathcal{J}_{ij}\geq  0$ and $\mathcal{K}_{ij}= 0$ for all $i,j =1, \dots , N$. Then $\mathcal{Z}_N$ given in \eqref{Ex} has the Lee-Yang property.
\end{proposition}
As mentioned above, we prove that $\mathcal{Z}_N$ has the Lee-Yang property for all $\theta>0$ and $\mathcal{K}_{ij}\in \mathds{R}$ whenever each positive $\mathcal{J}_{ij}$ is greater than some explicit bound defined by $\theta$ and $\mathcal{K}_{ij}$ with the same $i,j$. 

\subsection{The statement}

We begin by introducing the following functions  
\begin{gather}
 \label{E}
\kappa (v) = \ln (v + \sqrt{v^2-1 }), \quad v\geq 1 , \\ \lambda (w) = \ln( 1+ 2e^{-w}+ 2e^{-w} \sqrt{e^w +1}), \ w> 0. \nonumber
\end{gather}
Note that $\kappa(1) =0$ and $\cosh \kappa (v) = v$. Next, we set
\begin{equation}
\label{EA}
\Lambda (w) = \left\{\begin{array}{cc}
\kappa (e^{-w}) , \quad   & {\rm for} \quad w\leq 0, \\
    \lambda (w), \quad & {\rm for} \quad w> 0.
\end{array} \right.
\end{equation}
Now for $\chi_\theta$ as in \eqref{Ex1}, we define
\begin{eqnarray}
\label{E2}
& & F_{N} (x_1, \dots , x_{N}) \\ \nonumber & & \quad  =   \int_{\mathds{R}^{N}} \exp\left(\sum_{1\leq i<j\leq N } \left[\mathcal{J}_{ij} S_i S_j + \mathcal{K}_{ij} S_i^2 S^2_j \right]  + \sum_{i=1}^{N} x_i S_i\right) \prod_{i=1}^N \chi_\theta (d S_i) \\ \nonumber
& & \quad  =  \exp\left(\sum_{1\leq i<j\leq N } \left[\mathcal{J}_{ij} D_i D_j + \mathcal{K}_{ij}  D_i^2 D^2_j \right] \right) \prod_{i=1}^N \phi_\theta (x_i)    , \quad D_i = \frac{\partial}{\partial x_i},
\end{eqnarray}
where the interaction matrices $(\mathcal{J}_{ij})$ and $(\mathcal{K}_{ij})$ are the same as in \eqref{Ex}.  
The next statement is the main result of this work.
\begin{theorem}
 \label{1tm}
$F_N$ given in \eqref{E2} is in $\mathcal{F}_N$, and thus the partition function $\mathcal{Z}_N (x) = F_N (x, \cdots , x)$, has the Lee-Yang property if the following holds:
\begin{itemize}
    \item[(a)] For $\theta \in (0,1]$: $\mathcal{J}_{ij} \geq \Lambda (\mathcal{K}_{ij})$ for each pair $i,j$.
    \item [(b)] For $\theta >1$: $N=2M$, and there exists a division, $\{1, \dots , 2M\}=\cup_{k=1}^M \vartheta_k$, into disjoint two-element subsets (dimers)  $\vartheta_k = \{i_k, j_k\}$ such that $\mathcal{J}_{i j}  \geq  \Lambda(\mathcal{K}_{ij}) + \epsilon_{ij} \kappa (\theta^2)$ for each pair $i,j$. Here $\kappa$ and $\Lambda$ are as in \eqref{E} and \eqref{EA}, and $\epsilon_{ij} =1$ if $\{i,j\} = \vartheta_k$ for some $k$, and $\epsilon_{ij}=0$ otherwise.
\end{itemize}
 \end{theorem}
Note that in items (a) and (b) of Theorem \ref{1tm}, each $\mathcal{K}_{ij}$ can be positive or negative. This theorem describes also the Blume-Capel model, studied in \cite{KH}. It corresponds to $\mathcal{K}_{ij}=0$, i.e., $\Lambda(\mathcal{K}_{ij})=0 $; see \eqref{EA}.    
The condition $N=2M$ imposed in item (b) is not burdensome. In applications, it suffices to establish the Lee-Yang property of the partition function only for even $N$. The assumed positivity of $(\mathcal{J}_{ij})$ means that the graph on $\{1, \dots , 2M\}$ with the edge set $E_N =\{ \{i,j\}: \mathcal{J}_{ij} \geq \Lambda(\mathcal{K}_{ij}) +  \kappa (\theta^2)\}$ is perfectly matchable (or admits `dimer covering' in the terminology of \cite{HL}). The latter property is typical for models defined on graphs, such as $\mathds{Z}^d$, with sufficiently strong nearest-neighbor pair interactions. Similar conditions were imposed in \cite{KH}, where they are discussed in more detail.

\subsection{The proof}
As was already mentioned, an obstacle for us in proving the Lee-Yang property of $\mathcal{Z}_N$ is the quadrupole interaction term in the exponent in \eqref{E2}. 
We begin by deriving useful formulas that will allow us to exclude it.

For a fixed $\varkappa \in \mathds{R}$, let us find the smallest $\tau \geq 0$ such that the following equation 
\begin{equation}
\label{F}
e^{\tau \sigma + \varkappa \sigma^2} = (1+ a\sigma) (1+ b\sigma) , \qquad \sigma = \pm 1,
\end{equation}
has solutions $a,b\geq 0$. Taking into account the two possibilities for $\sigma$, we transform \eqref{F} into the following system of equations
\begin{equation*}
 \cosh \tau e^\varkappa = 1+ ab \quad {\rm and} \quad \sinh \tau e^\varkappa = a+b, 
\end{equation*}
with $a$ and $b$ as unknown quantities. Its solution is
\begin{eqnarray}
 \label{F2}
a & = & \frac{1}{2}\left[\sinh \tau e^\varkappa  + \sqrt{\left(\cosh \tau e^{\varkappa} -2\right)^2 - e^{2\varkappa} }\right], \\ \nonumber b &=& \frac{1}{2}\left[\sinh \tau e^\varkappa  - \sqrt{\left(\cosh \tau e^{\varkappa} -2\right)^2 - e^{2\varkappa} }\right]  .
\end{eqnarray}
For $\varkappa \leq 0$, the positivity of $b$ implies $\cosh \tau e^\varkappa \geq 1$. The least $\tau\geq 0$ that satisfies this demand, as well as the following one 
\begin{eqnarray}
 \label{F2a}
 |\cosh \tau e^{\varkappa} -2| \geq e^\varkappa,
\end{eqnarray}
-- since $a$ and $b$ should be real -- corresponds to 
\begin{equation}
\label{F3}
    \cosh \tau = e^{-\varkappa}, \quad {\rm which} \ \ {\rm yields} \ \ a = \sqrt{1- e^{2\varkappa}}, \ \ b=0.
\end{equation}
In particular, $a=b=0$ for $\varkappa =0$. For $\varkappa>0$, the condition in \eqref{F2a} takes the form $\cosh \tau \geq 2 e^{-\varkappa}+1$, which also guaranties $b\geq 0$. The least such $\tau\geq 0$ corresponds to the equality, which yields
\begin{equation}
 \label{F4}
e^\tau = 1+ 2e^{-\varkappa}+ 2e^{-\varkappa} \sqrt{e^\varkappa +1}, \quad   a = b = \sqrt{e^\varkappa +1}.
\end{equation}
Both versions \eqref{F3} and \eqref{F4} can be summarized in the following
\begin{equation}
 \label{F5}
 \tau = \Lambda (\varkappa);
\end{equation}
see \eqref{EA}. Now we define, cf. \eqref{F3} and \eqref{F4},
\begin{gather}
\label{F6}
\mathcal{A}_{ij} = \left\{\begin{array}{cc}
   \sqrt{1- e^{2\mathcal{K}_{ij}}}  & {\rm for} \ \mathcal{K}_{ij}\leq 0, \\[.2cm]
   \sqrt{1+ e^{\mathcal{K}_{ij}}}  & {\rm for} \ \mathcal{K}_{ij}> 0,   
\end{array} \right.
\end{gather}
and $\mathcal{B}_{ij} = 0$ for $\mathcal{K}_{ij}\leq 0$, $\mathcal{B}_{ij} = \mathcal{A}_{ij}$  for $\mathcal{K}_{ij}>0$. By means of \eqref{F2}, we can write
\begin{eqnarray}
\label{F7}
& & \exp\left(\mathcal{J}_{ij} S_iS_j + \mathcal{K}_{ij} S^2_iS^2_j  \right)\\ \nonumber & & \quad  = (1+\mathcal{A}_{ij} S_iS_j)(1 + \mathcal{B}_{ij} S_iS_j) \exp\left([\widetilde{\mathcal{J}}_{ij} + \epsilon_{ij}\tilde{\kappa} (\theta^2)] S_iS_j \right), \\ \nonumber
& & \quad  \widetilde{\mathcal{J}}_{ij} = \mathcal{J}_{ij} - \Lambda (\mathcal{K}_{ij}) - \epsilon_{ij}\tilde{\kappa} (\theta^2), \quad \tilde{\kappa} (\theta^2) = \left\{ \begin{array}{cc}
    \kappa (\theta^2) & {\rm for} \ \theta \geq 1, \\
 0    &{\rm otherwise}. 
\end{array}\right.
\end{eqnarray}
Then by \eqref{E2} and \eqref{F7}, we can decompose
\begin{eqnarray}
\label{F8}
F_N (x_1, \dots, x_N) & = & \prod_{1\leq i< j\leq N} (1+ \mathcal{A}_{ij} D_iD_j)(1 + \mathcal{B}_{ij} D_iD_j) G_N (x_1, \dots, x_N), \qquad  \\ \nonumber G_N (x_1, \dots, x_N) & = & \exp\left( \sum_{1\leq i< j\leq N}[\widetilde{\mathcal{J}}_{ij} + \epsilon_{ij}\tilde{\kappa} (\theta^2)] D_iD_j  \right) \prod_{i=1}^N \phi_\theta (x_i).
\end{eqnarray}
According to our assumptions, $\widetilde{\mathcal{J}}_{ij}\geq 0$ for all pairs $i,j$. The same is true for  
$\mathcal{A}_{ij}$ and $\mathcal{B}_{ij}$; see \eqref{F6}. For $\theta \leq 1$, we have $\phi_\theta \in \mathcal{F}_1$. Then the proof of claim (a) of Theorem \ref{1tm} follows by \eqref{F8} and Proposition \ref{0pn}. 

In the proof of claim (b), we mostly follow the way elaborated in \cite{KH}. Recall that we assume $N=2M$ and the existence of the decomposition $\{1, 2, \dots,  N\} = \cup_{k=1}^M \vartheta_k$ with the properties mentioned herein. Consequently, we further decompose $G_N$ in the following way  
\begin{eqnarray}
\label{F9}
G_N (x_1, \dots, x_N) & = & \exp\left( \sum_{1\leq i< j\leq N}\widetilde{\mathcal{J}}_{ij} D_i D_j\right) H_N (x_1, \dots, x_N), \\ \nonumber
H_N(x_1, \dots, x_N) & = & \exp\left(\sum_{k=1}^M  \kappa (\theta^2) D_{i_k} D_{j_k} \right) \prod_{i=1}^N \phi_\theta (x_i) = \prod_{k=1}^M \Phi (x_{i_k}, x_{j_k}).
\end{eqnarray}
Here
\begin{equation*}
 \Phi(x,y) = \exp\left(\kappa (\theta^2) D_x D_y \right) \phi_\theta (x) \phi_\theta (y).
\end{equation*}
To prove claim (b), we show that $\Phi \in \mathcal{F}_2$ even though $\phi_\theta$ is not in $\mathcal{F}_1$ for $\theta >1$; see \eqref{Ex2}. To this end, we write
\begin{eqnarray*}
    \Phi (x,y)  & = & 4\bigg{(} \theta^2 +  \theta( \cosh x +\cosh y)  \\ & + & [\cosh (\kappa(\theta^2) D_x D_y) + \sinh (\kappa(\theta^2) D_x D_y) ] \cosh x \cosh y\bigg{)},
\end{eqnarray*}
and take into account that
\begin{gather*}
 \cosh (\kappa(\theta^2) D_x D_y)  \cosh x \cosh y = \cosh (\kappa(\theta^2)) \cosh x \cosh y ,\\  \sinh (\kappa(\theta^2) D_x D_y)  \cosh x \cosh y = \sinh (\kappa(\theta^2)) \sinh x \sinh y,
\end{gather*}
as well as that
$\theta^2 = \cosh \kappa (\theta^2)$, see \eqref{E}. Thereafter, we arrive at the following formula 
\begin{eqnarray}
\label{F11}
 \Phi (x,y)  & = & 4\bigg{(}2 \theta \cosh u \cosh v + e^{\kappa(\theta^2)}\cosh^2 u + e^{-\kappa(\theta^2)}\cosh^2 v \bigg{)}\\ & = & 4 e^{\kappa(\theta^2)} \varphi_{+} (x,y)  \varphi_{-} (x,y), \quad u = \frac{x+y}{2}, \ \ v = \frac{x-y}{2}, \nonumber
\end{eqnarray}
where 
\begin{equation}
\label{F12}
\varphi_{\pm} (x, y) = \cosh \frac{x+y}{2} +\omega_{\pm}\cosh \frac{x-y}{2}, 
\end{equation}
and
\begin{equation*}
 \omega_{\pm} = e^{-\kappa(\theta^2)} \left[\theta \pm \sqrt{\theta^2-1}\right] = \exp\left[- \kappa(\theta^2) \pm \kappa(\theta) \right].
\end{equation*}
For $\theta>1$, one sees that $\kappa (\theta^2) > \kappa (\theta)$, which means that both $\omega_{\pm}$ are in $(0,1)$.
Consequently,
\[
\nu_{\pm} := \frac{1- \omega_{\pm}}{1+\omega_{\pm}} \in (0,1),
\]
and hence, see \eqref{F12},
\begin{equation*}
 \varphi_{\pm} (x,y) = (1+\omega_{\pm})\left[\cosh \frac{x}{2} \cosh \frac{y}{2} + \nu_{\pm} \sinh \frac{x}{2} \sinh \frac{y}{2}\right].
\end{equation*}
By \eqref{F11}, $\Phi \in \mathcal{F}_2$ if both $\varphi_{\pm}$ have this property. Clearly, both $\cosh(x/2)$ and $\sinh(x/2)$ can be approximated by the sequences of stable polynomials $\{c_n (x)\}$ and $\{s_n(x)\}$, respectively. Then  $\varphi_{\pm} \in \mathcal{F}_2$ if 
\begin{equation*}
 \gamma_n (x,y) = c_n (x) c_n (y)+ \nu s_n (x) s_n (y), \quad \nu \in (0,1),
\end{equation*}
is a stable polynomial for all sufficiently large $n$, i.e., it does not vanish for all $\Re x>0$, $\Re y>0$ and such $n$. The latter is equivalent to 
\begin{equation}
\label{E14}
\frac{c_n(x)}{s_n(x)} \neq - \nu \frac{s_n(y)}{c_n(y)}, \qquad \Re x >0, \  \Re y >0.
\end{equation}
For $\Re y>0$, we set
\begin{eqnarray}
\label{E15}
\frac{s_n(y)}{c_n(y)} = \frac{p_n (y) -1 }{p_n(y) + 1}=\rho_n e^{i\alpha_n}, \quad \rho_n >0.
\end{eqnarray}
Here $p_n (y) = (c_n (y) + s_n (y))/ (c_n (y) - s_n (y))\to e^y$ for $\Re y>0$. By \eqref{E15}, it follows that 
\begin{equation}
\label{E16}
p_n(y) = \frac{1+ \rho_n e^{i\alpha_n}}{1- \rho_n e^{i\alpha_n}},\ \ \ {\rm hence}\ \ \ |p_n(y)|^2 = \frac{1+\rho_n^2 + 2 \rho_n \cos \alpha_n }{1+\rho_n^2 - 2 \rho_n \cos \alpha_n}. 
\end{equation}
Assume $\gamma_n (x,y) = 0$, which corresponds to the equality in \eqref{E14}, and hence, cf. \eqref{E15}, to the following
\[
\frac{c_n(x)}{s_n(x)} = - \nu \rho_n e^{i\alpha_n}.
\]
Similarly to \eqref{E16} we then get
\begin{equation}
 \label{E17}
 |p_n(x)|^2 = \frac{1+ \nu^2\rho_n^2 - 2 \nu \rho_n \cos \alpha_n }{1+ \nu^2\rho_n^2 + 2 \nu \rho_n \cos \alpha_n}.
\end{equation}
By construction, $|p_n(x)|^2 \to e^{2 \Re x}$ and $|p_n(y)|^2 \to e^{2 \Re y}$, point-wise on $\Re x>0$, $\Re y>0$. Hence, the right-hand sides of \eqref{E16} and \eqref{E17} are both strictly greater than one 
whenever $n>\bar{n}$ for some threshold $\bar{n}$ dependent on $x$ and $y$. Clearly, this is not possible for any value of $\cos \alpha_n$. The obtained contradiction implies \eqref{E14}, which completes the proof.

\section{The Blume-Emery-Griffiths Model}
 
The Blume-Emery-Griffiths (BEG) model was introduced
in \cite{BEG} to describe the $\lambda$-transition and phase separation
in He3--He4 mixtures. In general form, its Hamiltonian reads as follows
\begin{equation}
 \label{6}
 H_N = - \sum_{1\leq i< j\leq N} (J_{ij} S_i S_j + K_{ij} S_i^2 S_j^2) + \Delta \sum_{i=1}^N S_i^2 - \sum_{i=1}^N h S_i.
\end{equation}
Here, each spin $S_i$ takes values $0, \pm 1$, $J_{ij}$ and $K_{ij}$ are bilinear and biquadratic (quadrupole) interaction intensities, and $\Delta$ is a crystal anisotropy field.   
Typically \cite{BEG,Nihat,Wu}, the spins are indexed by the vertices of a graph, e.g.,  a crystal $\mathds{Z}^d$, in which case the (nearest-neighbor) interaction intensities $J$ and $K$ are assigned to the graph edges.   

Along with the aforementioned physical systems, this model is used in the theory of many other substances and phenomena; see \cite{Nihat,Wu} and the publications quoted therein. It
has a variety of significant properties that are essentially different from those of the classical $S=\pm1$ Ising model. The difference is especially spectacular if the quadrupole intensities satisfy $K_{ij}\leq 0$ for all $i,j$; see \cite{Nihat,Lima,Wu}. Its partition function $Z^\beta_N$, cf. \eqref{3}, can be obtained from that in    
\eqref{Ex} by the formula $Z^\beta_N (h) = 2^{-\beta N \Delta }\mathcal{Z}_N (\beta h)$, in which the corresponding parameters and those of \eqref{6} are related to each other as follows
\begin{equation}
\label{A}
 \mathcal{J}_{ij} = \beta J_{ij}, \quad \mathcal{K}_{ij} = \beta K_{ij}, \quad \theta = e^{\beta \Delta}/2.
\end{equation}
According to Theorem \ref{1tm}, in the case of nearest-neighbor interactions, the partition function $Z^\beta_N (h)$ of the BEG model (with $N=2M$)  has the Lee-Yang property if the model parameters and $\beta$ satisfy
\begin{equation}
\label{A1}
\beta J \geq \Lambda (\beta K) + \tilde{\kappa} (e^{2\beta \Delta}/4);
\end{equation}
see \eqref{EA}, \eqref{F7} and \eqref{A}. Here, $K$ can be positive or negative. For $K=0$, this condition coincides with the corresponding condition derived in \cite{KH}. For $K<0$ and $\beta > \Delta^{-1} \ln 2$, \eqref{A1} turns into the following bound, cf. \eqref{E} and \eqref{EA},
\begin{gather}
 \label{A2}
 \beta J \geq \beta ( |K| + 2 \Delta) - \varepsilon (\beta),
 \end{gather}
 where
\begin{eqnarray*}
\varepsilon (\beta) & = & \ln 4 - \ln \left(1+ \sqrt{1-e^{-2\beta |K|}} \right) - \ln \left(1+ \sqrt{1-e^{-4\beta\Delta}/16} \right) \\ & = & \frac{1}{4} \bigg{(}e^{-2\beta |K|} + \frac{1}{16} e^{-4\beta \Delta}\bigg{)} + o \left(\max\{e^{-2\beta |K|}; e^{-4\beta \Delta}\} \right), \quad \beta \to + \infty.  
\end{eqnarray*}
Therefore, the partition function of the BEG model $Z^\beta_N$ has the Lee-Yang property for all $\beta \geq \Delta^{-1} \ln 2$ whenever $J \geq |K| + 2\Delta$. For $\beta < \Delta^{-1} \ln 2$, the condition in \eqref{A1} takes the form
\begin{equation}
 \label{A3}
 J-|K| \geq  \beta^{-1}\ln \left(1+ \sqrt{1- e^{-2\beta |K|}} \right).
\end{equation}
The latter cannot be satisfied for all $\beta \in (0, \Delta^{-1} \ln 2)$ since the right-hand side is unbounded on this interval.

In the case $K>0$, the large $\beta$ behavior of $\Lambda (\beta K)$ is completely different: by \eqref{E} and \eqref{EA}, one sees that $\Lambda (\beta K) \leq 4 e^{-\beta K/2}$. Therefore, instead of \eqref{A2} one gets the following condition
\begin{equation*}
 \beta J \geq 2\beta\Delta - \ln 4 + \ln \left(1+ \sqrt{1-e^{-4\beta\Delta}/16} \right) + 4 e^{-\beta K/2}. 
\end{equation*}
For $\beta < \Delta^{-1}\ln 2$, instead of \eqref{A3} one obtains
\begin{equation*}
J  \geq \beta^{-1} \ln \left(1 + 2e^{-\beta K} + 2 e^{-\beta K} \sqrt{1+ e^{\beta K}} \right) ,
\end{equation*}
which is also unbounded on $\beta \in (0, \Delta^{-1} \ln 2)$.







\end{document}